\documentclass{elsart}

 \usepackage{graphicx}

 \usepackage{epsfig}

\usepackage{amssymb}

\begin{document}

\begin{frontmatter}

\title{Current perpendicular to plane Giant Magnetoresistance (GMR) in laminated nanostructures}

\author[aff1,aff2]{A. Vedyayev}
%
\author[aff1]{I. Zhukov\corauthref{cor1}}
\corauth[cor1]{Tel: 007 (095) 939-47-87   ; fax: 007 (095)
939-47-87 } \ead{ilya@magn.ru}
\author[aff2]{B. Dieny}
\address[aff1]{Magnetism Department,
Faculty of Physics, M. V. Lomonosov Moscow State University,
Vorobievy gory, 119992 Moscow (Russia)}
\address[aff2]{"SPINTEC-Unite de Recherche Assosiee
CEA/DSM $\&$ CNRS/SPM-STIC" CEA Grenoble, 17 rue des Martyres,
38054 Grenoble Cedex (France)}

\begin{abstract}
We theoretically studied spin dependent electron transport
perpendicular-to-plain (CPP) in magnetic laminated multilayered
structures by using Kubo formalism. We took into account not only
bulk scattering, but the interface resistance due to both specular
and diffuse reflection and also spin conserving and spin-flip
processes. It was shown that spin-flip scattering at interfaces
substantially reduces the value of GMR. This can explain the
experimental observations that the CPP GMR ratio for laminated
structures only slightly increases as compared to non-laminated
ones despite lamination induces a significant increase in CPP
resistance.
\end{abstract}

\end{frontmatter}

We considered the spin dependent electron transport
perpendicular-to-plain (CPP) in magnetic multilayered structures
of the type (N/F)$_n$(N$^{\mathrm{spacer}}$)(F/N)$_m$, where F is
ferromagnetic, N is nonmagnetic metal, $n$ and $m$ denote the
number of bilayer repeats. The thickness of N layers in the
(N/F)$_n$ is small enough ($\sim5$\AA) so that the adjacent F
layers are coupled ferromagnetically and the thickness of
(N$^{\mathrm{spacer}})$ layer is larger ($\sim20$\AA) so that the
total magnetization of (N/F)$_n$ and (F/N)$_m$ can be switched
from parallel to antiparallel orientation by an external magnetic
field.

We took into account that Fermi momentum of the majority spin
subband in Co practically coincides with Fermi momentum of Cu, and
for minority spin subband these momenta are quite different. So
for parallel orientation of the magnetizations in all layers the
majority spin channel has very low resistance, in contrast
minority spin electrons undergo fort reflection at Co/Cu
interfaces so that this channel has a low conductivity. For the
antiparallel orientation of magnetizations of (N/F)$_n$ and
(N/F)$_m$ stacks, both spin channels exhibit low conductivity. The
change of overall conductivity between parallel and antiparallel
magnetic configurations determines the CPP-GMR amplitude. In
contrast to the present study, we emphasis that in Valet and Fert
theory of CPP-GMR~\cite{Valet-Fert}, the difference of Fermi
momentum for different metals and different spin subbands was not
taken into account.

The spin-dependent current through the system has been calculated
in Kubo formalism using the same approach as in Ref.~\cite{Kane}. We
calculated the Green function $G_\kappa(z,z^\prime)$ for the
considered multilayer. It is the solution of the following
equation~\cite{VedyTree97}:
\begin{equation}
\left[\frac{\hbar^2}{2m}\left(\frac{\partial^2}{\partial
z^2}-\kappa^2\right)+ E_f - \Sigma(z)\right]
G_\kappa(z,z^\prime)\\=\delta(z-z^\prime),\label{EqG}
\end{equation}
where $m$ is the effective electron mass, $\hbar\kappa$ is the
electron momentum in $XY$ - plane of the film, and $z$ is the
direction perpendicular to the film plane, $\Sigma$ is the
coherent potential (CP). We include in the definition of the
coherent potential not only the usual contributions of the
spin-dependent electron scattering in the bulk of every layer
(which influences the imaginary part of the CP), but we also take
into account the difference in the positions of the bottoms of the
spin subbands in Cu and Co, which enters in the real part of the
CP. As a result $\Sigma(z)$ is step function of coordinate and
depends on electron spin and relative orientation of
magnetizations in the two magnetic laminated layers.

Besides specular reflection at interfaces, electron may undergo
diffuse scattering due to interfacial roughness. So we add to the
bulk CP interfacial contribution in the form
$\Sigma^{\mathrm{in}}(z_i)=V_i\delta(z-z_i)$, where $z_i$ is the
position of interface $i$. $\Sigma^{\mathrm{in}}(z_i)$ is
calculated from CPA equation. In calculation of the
$\Sigma^{\mathrm{in}}(z_i)$ we take into account spin conserving
and spin flip interfacial scattering. For that the CP the vertex
corrections were calculated within the same approximations. The
bulk vertex correction are taken into account by introducing an
effective electric field which has been calculated by solving the
equation of nondivergency of the current as in~\cite{Camblong}:
\begin{equation}
j^\uparrow(z) = \int \sigma^\uparrow(z,z')
E^\uparrow(z') dz' = const \label{JConst1}
\end{equation}
\begin{equation}
j^\downarrow(z) = \int \sigma^\downarrow(z,z')
E^\downarrow(z') dz' = const \label{JConst2}
\end{equation}

Fig. 1 shows the dependence of the GMR of the structure
(Cu(5\AA)Co(b))$_3$ Cu(30\AA)(Co(b)Cu(5\AA))$_3$ on the thickness
b of the Co layer. The parameters are given in the caption of fig.
1. These correspond to a case where the spin-conserving
interfacial scattering gives substantial contribution to the total
resistance of the system. The curve exhibits an oscillatory
behavior with relatively high values at resonances, when
approximately the condition $k_fb=2\pi n$ is fulfilled. In the
considered case of large diffuse scattering, the influence of spin
flip processes is relatively weak.  In fig. 2 a similar dependence
is shown in a case of relatively weak diffuse scattering at
interfaces. The value of GMR increases in comparison to the former
case, if there no spin-flip scattering.

In experiment \cite{Eid} for the laminated structure
Py(60\AA)/Cu(40\AA) [Co(10\AA) /Cu(5\AA)]$_3$ Cu(35\AA) the
reported value of the GMR is 40\%, meanwhile for the non-laminated
one the GMR value is 25\%. These values are close to the results
of our calculations. We came to the conclusion that lamination may
really enhance the GMR value due to the additional spin dependent
reflection at Cu/Co interfaces. However, the strong diffuse
interfacial scattering, especially the spin flip one diminishes
this effect. In conclusion, lamination of the CoCuCo structure
does not increase the GMR value due to the interplay of specular
reflection and diffuse scattering on interfaces.

The work was partly financially supported by the Russian fund for
fundamental research (Grant N 04-02-16688a).

\begin{figure}
\centering
\includegraphics[width=9cm]{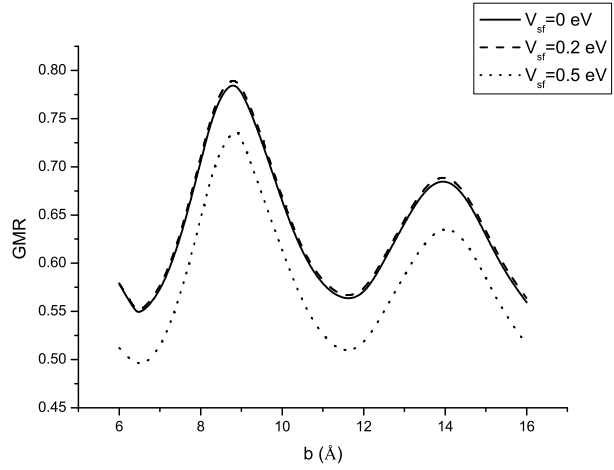}
\caption{The GMR dependence on Co thickness b for the structure
$(Cu(5\AA)Co(b))_3 Cu(30) (Co(b)Cu(5\AA))_3$ in the strong
interface scattering. Where $k_{f\,Co}^\uparrow=1\AA^{-1}$,
$k_{f\,Co}^\downarrow=0.6\AA^{-1}$,
$k_{f\,Cu}^\uparrow=k_{f\,Cu}^\downarrow=1\AA^{-1}$;
$l_{Co}^\uparrow=100\AA$, $l_{Co}^\downarrow=100\AA$,
$l_{Cu}^\uparrow=l_{Cu}^\downarrow=200\AA$, $k_{f\,i}$ are the
corresponding Fermi momentum, $l_i$ are the corresponding mean
free paths. $V_{Co/Cu}^\uparrow=1.1$ eV,
$V_{Co/Cu}^\downarrow=1.7$ eV are the scattering potentials on the
Co/Cu interfaces, x=0.5 is impurities concentration. For the
initial system Co(30\AA)/Cu(30\AA)/Co(30\AA) the GMR value is 0.51
for the same parameters. \label{GMRStrong}}
\end{figure}
\begin{figure}
\centering
\includegraphics[width=9cm]{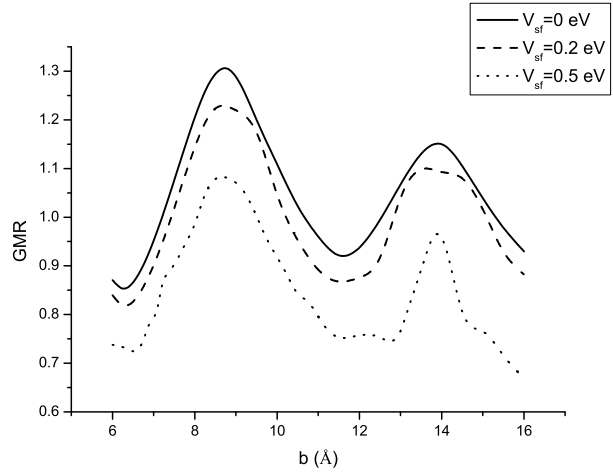}
\caption{Same dependence of GMR in the case of weak interfacial
scattering. $\mathrm{Im}\,V_{Co/Cu}^\uparrow=0.1$ eV.
$V_{Co/Cu}^\uparrow=0.1$ eV, $V_{Co/Cu}^\downarrow=0.3$ eV are the
scattering potentials on the Co/Cu interfaces, x=0.5 is impurities
concentration. For the initial system
Co(30\AA)/Cu(30\AA)/Co(30\AA) the GMR value is 0.54 for the same
parameters. \label{GMRWeak}}
\end{figure}

\end{document}